\documentclass[reprint,aps,twocolumn,groupedaddress]{revtex4-2}
\usepackage{graphicx}
\usepackage{dcolumn}
\usepackage{bm}
\usepackage{xfrac} 
\usepackage{amssymb,amsmath}

\begin{document}


\title{Towards Quantum Gates with Wide Operating Margins}


\author{Ryan J. Epstein}
\affiliation{Northrop Grumman Corporation, Aurora, Colorado, 80017, USA}

\date{\today}

\begin{abstract}
Scaling up quantum computing hardware is hindered by the narrow operating margins of current quantum components. Here, we introduce a composite qubit and gate scheme that achieves wide margins by use of transistor-like nonlinearities to suppress the effects of both ambient noise and control signal imperfections. This is accomplished by adiabatic deformation of subsystem codes based on anti-commuting two-body interactions. We focus on a resource-efficient variation that exploits biased noise and preserves bias under gate operation. As a proof of concept, we present simulations of a superconducting circuit that demonstrates core elements of the approach and discuss the challenges of experimental implementation.

\end{abstract}


\maketitle

\section{\label{intro}Introduction}

Developing quantum components that are robust to noise and control signal distortion would be a significant accelerant to the construction of a large-scale quantum computer. Operating margins quantify how much noise and distortion a component can tolerate before its performance drops below some threshold value. A given hardware-specific quantum computing architecture levies requirements on gate fidelities which flow down to requirements on gate operating margins. Traditional quantum gates have intrinsically narrow operating margins that further decrease as the required gate fidelity increases. In this paper, we propose a gate approach for which the margins can be orders of magnitude wider and have favorable scaling.

In the field of superconducting electronics, there has been important progress in qubit circuit designs with improved robustness to noise. Several years ago, the transmon \cite{tmonkoch, tmonschust} demonstrated a dramatic reduction of sensitivity to static charge noise by tuning the ratio of Josephson and capacitive energies. More recently, fluxonium circuit variants have achieved increased coherence times by decreasing charge and current matrix elements between qubit states and by decreasing flux sensitivity \cite{fluxonium, t1-fluxonium, heavy-fluxonium, long-fluxonium, fluxonium-gates, ms-fluxonium}. On the other hand, comparatively little progress has been made in demonstrating quantum gates with noise-resilience or wide operating margins. Indeed, in a thorough review of noise-protected superconducting quantum circuits \cite{gyenis}, Gyenis et al. note the dearth of noise-protected gate schemes. The concepts presented here will hopefully catalyze further work in this area. 

How would one go about designing qubits and gates with greater resilience to noise? Kitaev pioneered the idea of protecting quantum information by encoding it in the degenerate ground space of strongly interacting many-body quantum systems \cite{anyons}. He also developed gate schemes based on braiding of particles that have some intrinsic robustness. His studies of 1-dimensional systems that produce unpaired Majorana fermions \cite{kitaev-wire} has generated much theoretical and experimental follow-on effort and is a source of inspiration for this work. 

Our previous gate methods \cite{epstein} built off of Adiabatic Gate Teleportation \cite{agt}, simplifying the ingredients to local fields, two-body interactions, and single ancilla qubits. These simple encoding schemes do not protect against local noise, however, and require the ability to turn off physical interactions to high precision. The work presented here addresses both of those shortcomings and is a novel encoded generalization of the previous method \cite{epstein}. It is most similar to adiabatic topological quantum computing \cite{cesare, brun14, brun15}, making use of adiabatic code deformations that generate noncyclic holonomies \cite{sjoqvist}. There are also similarities to holonomic gates and error suppression with spin chains \cite{renes} and subsystem codes \cite{oreshkovPRA, marvian}. The combination of features in this work -- low-resource encoding, wide operating margins, and nearest-neighbor two-body interactions -- is novel and favorable for experimental realization.

For physical implementation, rather than engineering a material system with suitable topological properties, we design superconducting circuits built from microscale components that emulate strongly coupled qubits. Arrays of superconducting rhombus circuits are also being explored for such purposes \cite{doucot, gladchenko, bell, rhombus-gates}. Our approach is similar but, while the rhombi require tuning of offset charge using gate voltages, we have found a design that avoids offset charge sensitivity. We further provide evidence that it is possible to strongly suppress sensitivity to all sources of low-frequency flux and charge noise, from both control signals and the surrounding material environment, throughout gate operation.

One of the most important features of the classical transistor, that leads to robust digital logic operation, is its highly nonlinear current-voltage characteristic. Here we show that our quantum \textsc{cnot} gate needs two types of couplings, XX and ZZ, and that both can be made to have transistor-like robustness due to the anti-commuting terms in our encoding. The proposed gates have two important properties: 1) Gate fidelity does not depend on the area (amplitude integrated over time) of any control pulse. Pulses merely need to overlap in time to maintain energy gaps. 2) No physical coupling strength needs to be set with high accuracy. When physical couplings are approximately turned off, encoded couplings are strongly turned off. When couplings are on, they merely need to be strong for adiabatic protection. 

\section{Pauli Model}
Our abstract gate model is based on two-level systems (i.e. qubits) and two-body interactions composed of tensor products of the Pauli operators X and Z. Logical qubits are encoded in the doubly degenerate ground space of XX-coupled qubit chains. Noise-resilient two-qubit gates are enacted through adiabatically evolved sequences of local fields and couplings between multiple chains. In isolation, a chain is described by the transverse field Ising model, with strong nearest-neighbor XX coupling terms protecting against local Z field noise. For isotropic qubits, the Ising chain does not protect against X fields which split the ground states linearly in X field strength. In our case, we make use of qubits that are anisotropic in that X fields are negligible; the remaining Z noise anti-commutes with the XX couplings \cite{imarvian}. Kitaev showed how the Ising model is equivalent to a Majorana fermion chain under a nonlocal transformation \cite{kitaev-tfim}. Here, we remain in the qubit picture but go beyond the Ising model to demonstrate robust gates using only local fields and nearest-neighbor two-body anti-commuting interactions.  In Section~\ref{circuits}, we also present a superconducting circuit which achieves both strong XX interactions and strongly suppressed X fields. This creates the desired anisotropic qubit system and preserves robustness to local noise, in contrast to \cite{nori}.

\begin{figure}
	\includegraphics{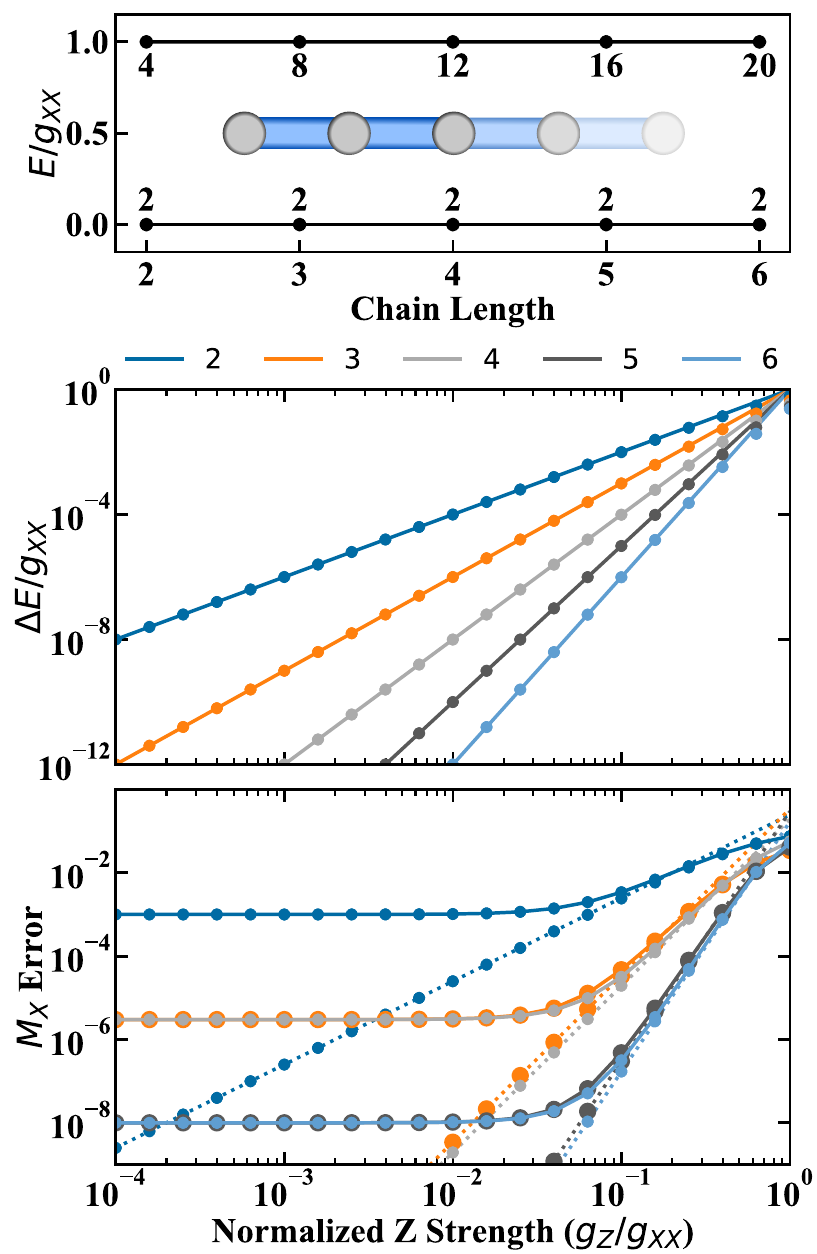}
	\caption{\label{fig:chain} Numerical simulation of XX-coupled chains. (top) Low-energy spectrum of the XX-coupled qubit chains as a function of chain length. Degeneracy is listed by each by data point. Inset: diagram of the chain. Grey discs are qubits, blue bonds are XX couplings, and color fade indicates that chain length is varied. (middle) Splitting of the ground state as a function of normalized Z field strength for different chain lengths. Solid lines are plots of $(g_Z/g_{XX})^L$ where $L$ is the chain length. (bottom) Error probability for measuring and decoding the logical X state for different chain lengths. Solid and dotted lines correspond to a single-qubit X measurement error of $10^{-3}$ and perfect measurement respectively. Dotted lines are fits to $c(g_Z/g_{XX})^b$, where $c$ is a fit parameter, and $b$ is 2, 4, 4, 6, and 6 for chain lengths of 2 through 6 respectively.}
\end{figure}

Figure \ref{fig:chain} provides a summary of the relevant properties of XX-coupled qubit chains. The top panel shows the low-energy spectrum. For increasing chain length, the ground state remains doubly degenerate, the energy gap is constant, and the excited state degeneracy at the gap increases linearly. 

It is beneficial to have an energy gap that is significantly larger than the thermal energy. While physical bit flips are assumed to be suppressed by the physical (superconducting) components, thermal excitation out of the ground space, followed by subsequent relaxation, can cause logical bit flips. In the large gap regime, however, the rate of thermal excitation out of the ground space is exponentially suppressed. Detailed balance implies that the rate of excitation from a ground state to a state $E$ above is slower than the rate of the reverse process by the factor $\exp(E/k_BT)$, where $k_B$ is the Boltzmann constant and $T$ is temperature. For $E/h=5$~GHz and $T = 40$~mK, the factor is roughly 400; doubling $E/h$ to 10~GHz increases the factor to $1.6\times10^5$. Such suppression of thermal excitation rates would relax requirements on the quality factor of dielectrics in a superconducting circuit for a given target bit flip rate. This has the potential to open up design space from single layer aluminum structures with minimal surrounding dielectric to multi-layer structures with cross-overs, parallel plate capacitors, etc., using standard dielectric layers.

The XX chain is designed to suppress the effects of Z fields. Figure~\ref{fig:chain}(middle) shows how the ground space splits with Z fields of the same strength applied to all qubits. The splitting decreases exponentially in chain length for a fixed Z field strength and is well approximated for small fields strengths by $(g_Z/g_{XX})^L$, where $g_Z$ is the Z field strength, $g_{XX}$ is the XX coupling strength and $L$ is the chain length (i.e. the number of qubits in the chain). This splitting causes dephasing of superpositions of logical 0 and 1 states and is exponentially suppressed in the chain length. In separate simulations, we have found that random Z fields produce the same scaling but with smaller prefactors (not shown).

\begin{figure*}
	\includegraphics{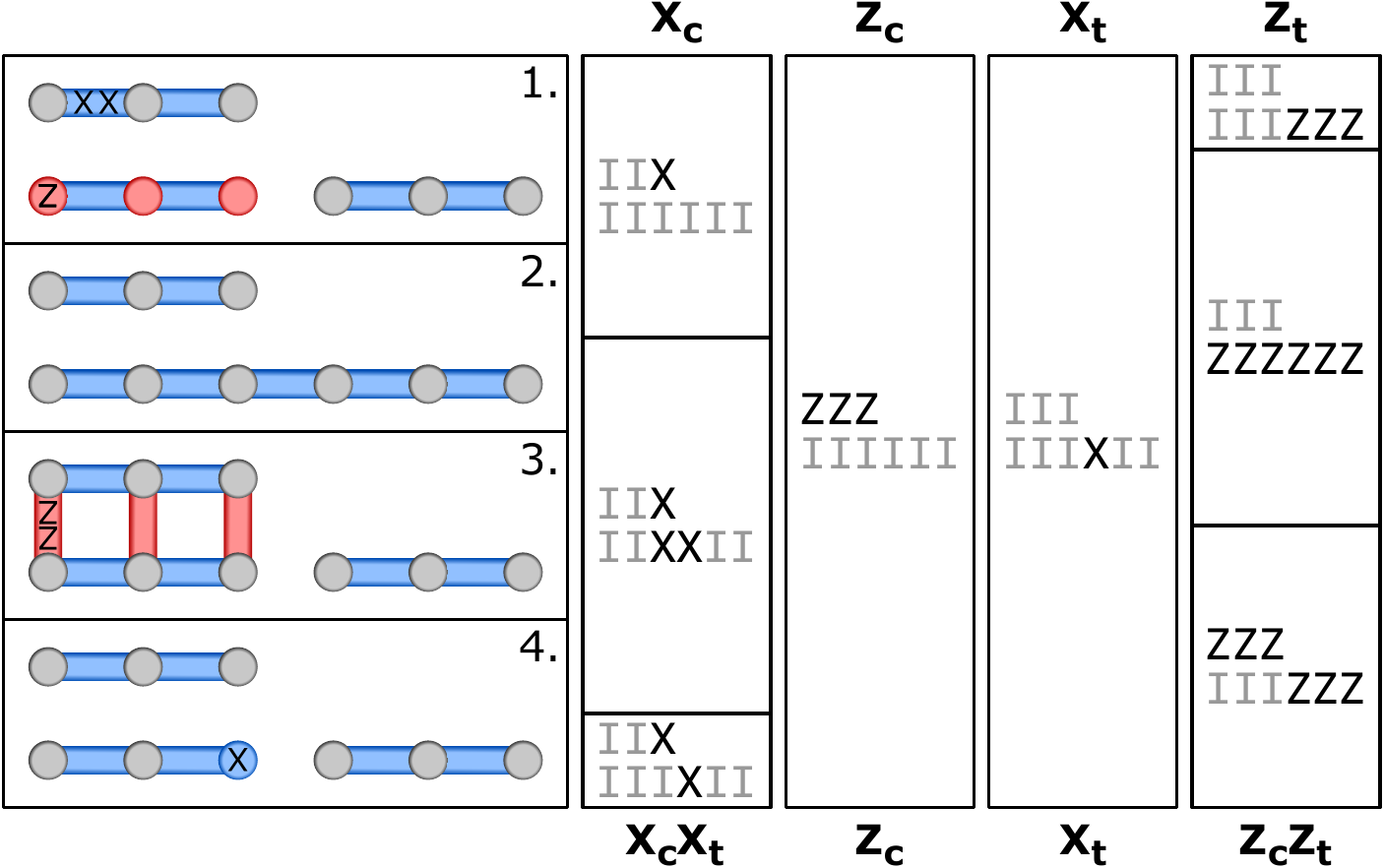}
	\caption{\label{fig:cnot-hams} Summary of protected \textsc{cnot} gate for length-3 encoded qubits.  Time flows from top to bottom. The gate proceeds via adiabatic interpolation between the four Hamiltonians shown pictorially on the left. Grey, red, and blue discs represent qubits with zero, Z, and X fields applied respectively. Red and blue bonds are ZZ and XX interactions respectively.  Logical operators for control (c) and target (t) qubits are tracked on the right. Boxes encompass regions where an associated operator commutes with the instantaneous Hamiltonian. Logical operators are labeled at top and bottom, showing that they transform as expected for a \textsc{cnot} gate. Note that, since the time-reversed sequence also produces a \textsc{cnot} gate, the logical ancilla need not return to its original state to perform a subsequent \textsc{cnot}.}
\end{figure*}

In addition to logical dephasing, Z fields induce phase flips upon single-qubit measurement in the X basis. (Since the Z fields commute with the logical Z operator, they do not affect logical Z measurement.) Logical X measurement is achieved by measuring X on each qubit in the chain and performing a decoding step. The ideal logical X eigenstates are $|++...+\rangle$ and $|--...-\rangle$. In the presence of Z fields and measurement error some outcomes will be flipped. Majority voting on the measurement outcomes is the decoding procedure used here. Figure~\ref{fig:chain}(bottom) shows the readout error with Z fields applied for perfect single qubit X measurement and for a measurement error probability of $10^{-3}$. Only a single round of measurement is performed. For small $g_Z/g_{XX}$, the logical X measurement error is dominated by the single qubit X measurement error, whereas for larger $g_Z/g_{XX}$ it is dominated by state ``corruption'' due to the Z fields. For both perfect and imperfect qubit measurements, the logical error decreases exponentially in the chain length rounded up to the nearest even length.

\section{Protected Gates}
We choose the universal gate set consisting of preparation and measurement in X and Z bases, a protected \textsc{cnot} gate, and an unprotected non-Clifford gate. This gate set is particularly well-suited for error correction schemes that involve multi-qubit X and Z parity measurements, which can be efficiently constructed from these gates. We also show that our \textsc{cnot} gate preserves noise bias.

\subsection{Gate Theory}
Figure~\ref{fig:cnot-hams} summarizes the protected \textsc{cnot} gate for encoded qubits of length 3. The left side of the figure depicts the Hamiltonian at four distinct times. The gate is performed by adiabatic interpolation between these Hamiltonians. The upper left chain is the logical control qubit, the lower left chain is a logical ancilla, and the lower right chain is the logical target qubit. On the right side of the figure, logical X and Z operator transformations are tracked. The gate operation works by the same principles as in Ref.~\cite{epstein}. Due to the encoding, a given logical operator has multiple equivalent forms. For example, the logical X operator on the control qubit is IIX, or equivalently IXI, or XII. An equivalent logical operator can be obtained by multiplying the original operator by any number of terms in the current Hamiltonian so long as the result still commutes with the Hamiltonian. For example, IXI is seen as a product of IIX and the Hamiltonian term IXX. In general the gate functions by sequential selection of logical operators based on their commutation with the instantaneous Hamiltonian.

Let us go through the evolution of two logical operators. First, consider the column labeled $Z_c$, the logical Z operator for the control qubit. It contains a single operator ZZZ;IIIIII inside a box that extends over the full vertical range. This means that the logical operator does not transform and that it commutes with all four Hamiltonians at the left. As another example, consider logical $X_c$. The operator starts as IIX;IIIIII at time step 1. At time step 2, an equivalent operator can be obtained by multiplying the operator by III;IIXXII, a term in Hamiltonian 2. Therefore, IIX;IIXXII is an equivalent logical operator that commutes with Hamiltonians 2-4. At time step 4, we can multiply IIX;IIXXII by Hamiltonian 4 term III;IIXIII to obtain IIX;IIIXII, which is clearly $X_cX_t$. The sequence of Hamiltonians thus induces the transformation $X_c \rightarrow X_cX_t$. By examining all four logical operator transformations, we see that the gate is indeed a \textsc{cnot}.

It is worth noting that a direct logical implementation of the scheme in Ref.~\cite{epstein} would entail promoting all physical couplings and fields to their logical counterparts. Here, physical X is equivalent to logical X but logical Z is a tensor product of each physical Z operator in the chain. Fortunately, the \textsc{cnot} gate presented here does not require such high weight operators. Physical Z fields and ZZ interactions suffice to control the evolution of the logical Pauli operators and to maintain error suppression of local Z fields \cite{jiang, marvian}. 

Note also that this gate works with the two-dimensional quantum compass (subsystem code) model \cite{doucot-compass, bacon-compass, bravyi}. It resembles lattice surgery \cite{horsman} except that gates are not performed directly on the physical qubits. Since we have identified a physical system that suppresses X fields, here we use a one-dimensional code which has quadratically lower qubit overhead. 

Let us briefly turn to state preparation and measurement. Logical Z state preparation can be achieved by applying strong Z fields on all qubits in an XX chain and allowing the system to relax to the ground state. Recall that with a lossy (dielectric) environment, this relaxation rate can be relatively fast even when the excitation rate out of the ground space is slow. Ramping off the Z fields prepares logical Z. The product of the signs of the Z fields determine whether logical 0 or logical 1 is prepared. Since the Z fields commute with logical Z, they do not need to be ramped off adiabatically. Alternatively, logical Z can be prepared based on the outcome of a (non-demolition) logical Z measurement, applying bit flips as needed.

Logical Z measurement can be achieved by measuring the Z operator of each qubit in the chain. The bit parity of the measurement corresponds to logical Z. This, however, requires all measurements to be correct. For a chain length of $N$ and single-qubit measurement error probability $p$, the logical measurement error is $Np$ to leading order. A alternative approach is to adiabatically apply strong Z fields to all but one qubit in a chain and then measure that qubit. The Z fields ``squeeze'' the quantum information to the measured qubit. Since Z fields commute with logical Z, they do not affect the measurement result. All other qubits ideally remain in state 0 so the bit parity is merely the bit state of the measured qubit. Therefore the  logical Z  measurement error only depends on the ``squeeze'' operation error and the single qubit measurement error.

Logical X measurement was discussed previously. Logical X preparation can be achieved by applying a strong X field to any single qubit in an XX chain and allowing the system to relax to the ground state. The gap is equal to the strength of X, which will determine the thermal ground state occupancy. Ramping off the X field will prepare logical X. Since the X field is equal to logical X and commutes with the Hamiltonian, it also need not be ramped off adiabatically. Alternatively, logical X can be prepared based on the outcome of a logical X measurement, where any errant bit can be flipped.

An unprotected non-Clifford gate can be achieved by simply applying a small X field for some duration. These gates would be as sensitive to control pulse amplitude and duration as traditional gates.

\subsection{Nonlinear Signal Transduction}
\begin{figure}
	\includegraphics{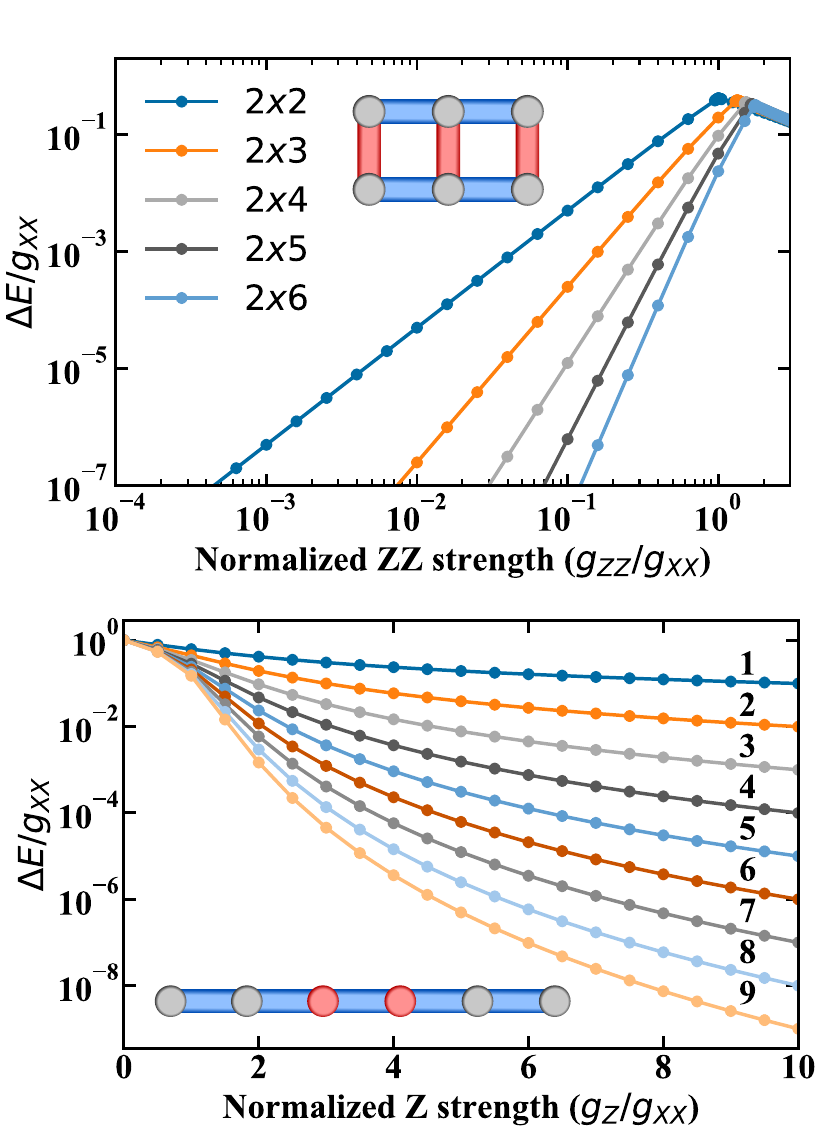}
	\caption{\label{fig:transistor} Transistor-like nonlinear response of energy levels for the protected \textsc{cnot} gate. (top) Two logical XX chains are coupled via ZZ interactions. Plotted is the logical coupling strength (the splitting between the logical state doublets) as a function of ZZ coupling strength. Each curve is for a different chain length. Inset is a diagram of the 2x3 case. (bottom) Two length-2 logical qubits XX-coupled via various numbers of intervening ancilla qubits with applied Z fields. Inset is a diagram for 2 intervening ancillae. Plotted is the logical coupling as a function of applied Z field strength. Number of ancillae for each curve is indicated on the plot. For both plots, the equal spacing between curves at a fixed $g_{ZZ}/g_{XX}$ (or $g_{Z}/g_{XX}$) implies exponential suppression of splitting in the number of (ancilla) qubits.}
\end{figure}

Now we explore the spectral properties of the protected \textsc{cnot} gate. We plot the dependence of \textit{logical} coupling strengths on \textit{physical} coupling (or field) strengths in Fig.~\ref{fig:transistor} for both XX and ZZ couplings. For nonzero physical ZZ couplings or weak Z field strengths, the 4-fold degenerate ground space splits into two doublets; we define the logical coupling strength as the energy splitting $\Delta E$ between the doublets. Importantly, the logical coupling strengths have a highly nonlinear dependence on the physical coupling strengths due to the anti-commuting Hamiltonian terms. This nonlinearity is crucial to engineering the operating margins of the gate. 

Figure~\ref{fig:transistor} (top) summarizes the ZZ coupling case. The logical coupling has a power law dependence on physical coupling at low ZZ strength with exponent equal to the chain length. An operating margin can be obtained from this plot. For example, suppose a given system architecture required a normalized logical coupling of less than $0.001\%$ for the ``off'' coupling condition. For chain lengths of 2 through 6, this flows down a requirement that physical coupling strengths be less than 0.45, 3.4, 9.5, 17.4, and 26.2\%, respectively. For a traditional non-encoded gate, the requirement on logical coupling strength \textit{is} the requirement on physical coupling strength. With this transistor-like nonlinearity, we have widened the ZZ coupling operating margin for the off condition by over 4 orders of magnitude for a chain length of 5.

For the``on'' condition, note how the logical coupling turns on as $g_{ZZ}/g_{XX}$ approaches and exceeds 1. The value of $g_{ZZ}/g_{XX}$ that gives the maximum energy gap in the coupled state depends on the chain length; the maximum gap decreases from 0.41 to 0.31 times $g_{XX}$ for chain lengths of 2 to 6. Note that, importantly, the ratio $g_{ZZ}/g_{XX}$ only needs to be coarsely adjusted to maintain a large gap; gate operation does not depend directly on the ``on'' coupling strengths, only that they are sufficiently strong to achieve adiabatic operation.

The \textsc{cnot} gate discussed here also requires a tunable XX coupling. While such a tunable coupling may be possible in a given physical implementation, we show how to construct one from fixed XX couplings and tunable Z fields. Figure~\ref{fig:transistor} (bottom) shows the splitting of the ground space for two chains coupled end-to-end as depicted in the inset. Some number of XX-coupled ancilla qubits are placed between the logical qubits; applying strong Z fields to them decouples the logical qubits. The plot shows that for large Z fields, the coupling decreases exponentially in the number of ancillae. As the Z fields are decreased, the coupling turns on and the energy gap reaches that of the XX chain. Like the ZZ coupling case, the nonlinear response leads to a widening of control signal operating margins. A full set of operating margins for the control pulses can be obtained from gate simulations.

\subsection{Gate Simulations}
\begin{figure}
	\includegraphics{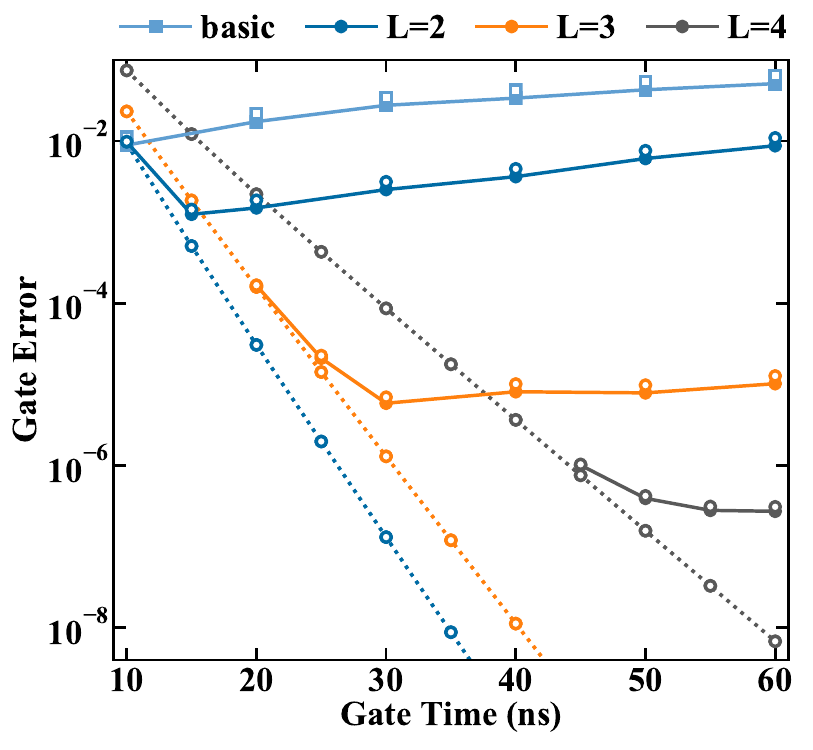}
	\caption{\label{fig:cnot-sim} Monte Carlo simulation of noise-resilient \textsc{cnot} gates. Average gate infidelity (solid symbols) and entanglement infidelity (open symbols) versus gate time for different chain lengths. Dotted lines are for noiseless control pulses. Solid lines are for noisy pulses and Z fields. Blue squares are for a basic two-qubit gate. The average gate infidelity \cite{pedersen} and entanglement infidelity \cite{nielsen, schumacher} are calculated per run as $1-(\mathrm{Tr}(\tilde{U}^\dagger \tilde{U}) + |\mathrm{Tr}(\tilde{U}^\dagger U)|^2)/d(d+1)$ and $1-|\mathrm{Tr}(\tilde{U}^\dagger U)|^2/d^2$, respectively, where $U$ is the ideal gate, and $\tilde{U}$ is the simulated gate in the computational subspace of dimension $d=4$. The data exhibits the expected ratio of entanglement infidelity to average gate infidelity of $(d+1)/d$ \cite{horodecki} when leakage error is negligible. All points with noise are averages of 100 Monte Carlo runs except for basic and $L=3$ points which are averages of 200 runs. Standard errors of the mean are smaller than the data symbols.}
\end{figure}

We now turn to numerical simulations of our \textsc{cnot} gate. Figure~\ref{fig:cnot-sim} contains a summary of simulations for different gate durations and chain lengths for XX couplings of 5~GHz. We plot both average gate infidelity and entanglement infidelity and refer to them both as gate error. Dotted lines show the case for no noise where gate error is dominated by nonadiabaticity. Control pulses are constructed from Gaussian error functions creating smooth turn-on and -off behavior. The pulse rise and fall times are a fixed fraction of the pulse widths. See the Appendix for further pulse details. The simulations show that as the gate time increases the gate error decreases exponentially. This is expected from going deeper into the adiabatic regime. The rate of exponential decay decreases as the chains get longer, however, for two reasons. First, the gap when the ZZ interactions are turned on gets smaller as the chains get longer as shown in Figure~\ref{fig:transistor}. Second, the degeneracy of states at the gap increases linearly with chain length as indicated in Fig.~\ref{fig:chain}. Both of these effects increase nonadiabitic excitation out of the ground space.

Solid lines are from simulations that include low-frequency classical noise. Noise is included on every Hamiltonian term and on Z fields applied independently to each qubit. The noise amplitude is 4 MHz rms and has been low-pass filtered with an exponential envelope of 0.25~GHz $1/e^2$ bandwidth. The strong filtering reduces noise at the gap frequency that causes direct excitation out of the ground space, which this gate scheme does not protect against. For Z and ZZ terms, the noise is added directly to the control pulses. For X and XX terms, the response to control pulses is assumed to be nonlinear due to e.g. the XX coupling scheme of Figure~\ref{fig:transistor} or the nonlinear response of tunneling (X fields) to barrier height. For simplicity, we take the pulse shape itself to be the nonlinear response function, so the noise is \textit{multiplied} by the pulses; noise is small when the pulse amplitude is small and is the full 4 MHz rms when the pulse is at full amplitude. ZZ couplings are assumed to be controlled by a single pulse and so receive the same noise instance. The same is true for Z fields applied to the logical ancilla in the gate operation. This worsens gate performance compared to uncorrelated noise (not shown).

For comparison, simulations of a basic \textsc{cnot} are also plotted for the same additive noise on the coupling and local Z fields. In this case, the gate is achieved by turning on a ZZ interaction with the same pulse shape as the protected gates. (Single qubit rotations that convert this gate to a \textsc{cnot} are assumed to be perfect.) The amplitude of the basic gate pulse is precisely adjusted for each gate time to minimize gate error. For the given noise power, the basic gate has error of roughly $10^{-2}$ at 10~ns, which increases with gate time. The length-2 protected case reduces gate error below that of the basic gate by roughly one order of magnitude. For length 3, the reduction is more than two orders of magnitude below that of length 2, and the length-4 case is over an order of magnitude below that of length 3. Like the noise-free case, the degree of error suppression depends on the minimum energy gap and the degeneracy at the gap.

In summary, the simulations reveal the remarkable robustness of the gate due to both the transistor-like nonlinearities creating wide margins for coupler turn-off, and the adiabatic gate approach enabling wide margins for coupler turn-on. More detailed simulations could include thermal effects, accurate nonlinear responses of X and XX terms, a physically motivated noise spectrum, and any Hamiltonian imperfections originating from a hardware-specific implementation.

\subsection{Noise Bias Preservation}
It has been proven by Guillaud et al. that a general class of gates on qubits cannot preserve noise bias \cite{guillaud}. Their no-go proof applies to gates that evolve as a generalized rotation from the identity. Fortunately, gates such as those described in \cite{epstein} and the protected \textsc{cnot} presented here, in addition to measurement-based gates \cite{childs}, are outside the scope of the no-go theorem.

It is very straightforward to see that our \textsc{cnot} gate preserves noise bias. Consider noisy Z fields applied to each physical qubit. As we saw earlier, such noisy fields can split the ground space, causing dephasing. However, by inspection every physical Z operator commutes with the evolving $\bar{Z}_t$ and $\bar{Z}_c$ throughout the entire \textsc{cnot} gate evolution. This proves that none of the physical Z operators pick up any logical X or Y component, thus preserving noise bias. The key feature of our gate is that the logical operators remain composed of Pauli operators of the same type (X, Y or Z) throughout the gate evolution. Logical operators for which the no-go applies rotate smoothly between linear combinations of Pauli types over the course of the gate evolution, thereby mixing noise components.

Note that the simulations summarized in Fig.~\ref{fig:cnot-sim} corroborate the claim of bias preservation. Since there is no protection for local X fields, any Z fields that would be mixed into X fields would induce large gate errors. The simulations show that no detectable mixing occurs.

\section{\label{circuits}Superconducting Circuit Implementation}

\begin{figure*}
	\includegraphics{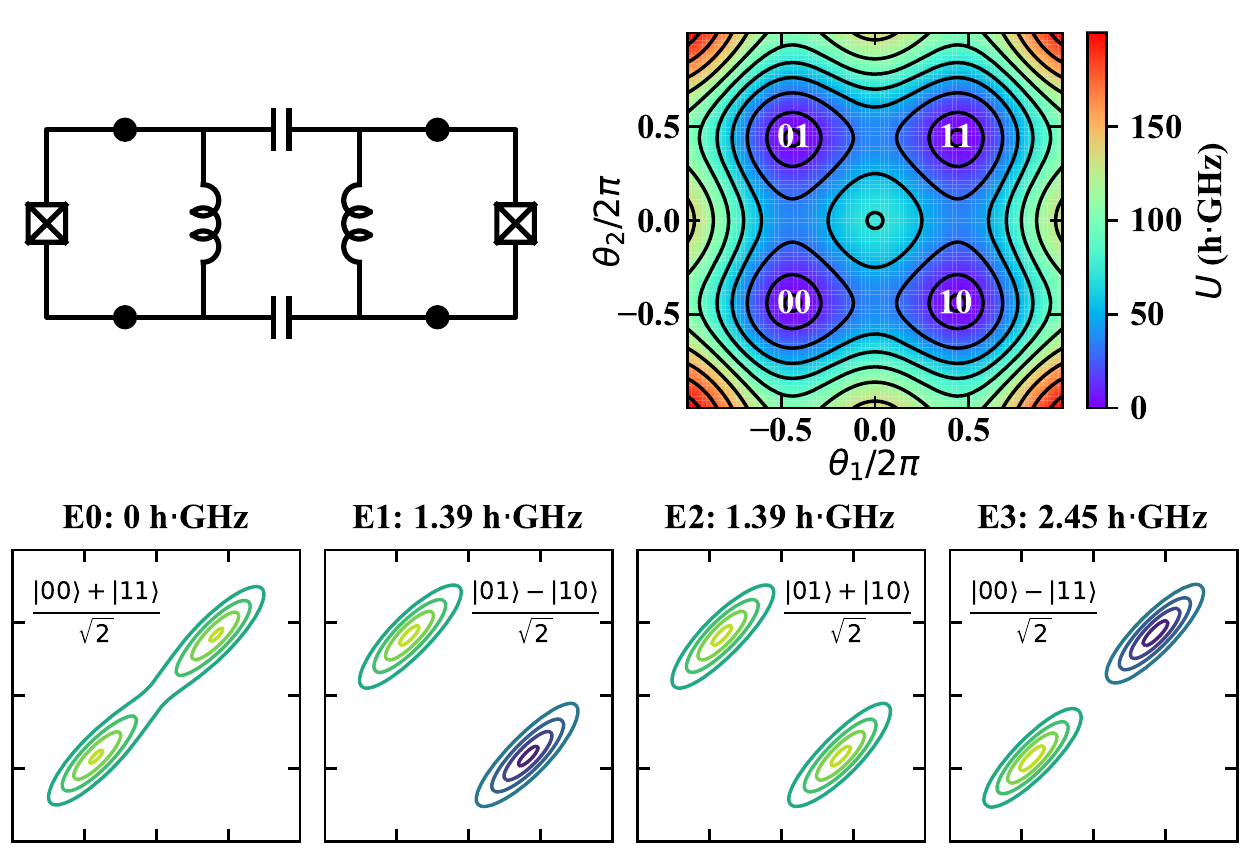}
	\caption{\label{fig:fluxonia} Two capacitively coupled fluxonium circuits.  (upper left) The circuit diagram. Circuit parameters are: junction critical current $I_c=50$~nA, inductance $L=50$~nH, coupling capacitance $C_c=150$~fF, capacitance to ground on each node $C_g=0.3$~fF, junction capacitance $C_J=0.75$~fF. (upper right) The potential $U$ as a function of phase differences across the two junctions. (lower) The first four lowest eigenstates with titles indicating eigenenergy. Axis ranges match the potential plot.}
\end{figure*}

In this section we present a proof-of-concept implementation of the above protected qubits and gates using superconducting circuits. The circuits are based on capacitive coupling of fluxonium-like qubits and exhibit all of the requirements for protected gates. The main limitation is that the circuit parameters in the regime of strong coupling are challenging to achieve experimentally. Nevertheless, it is a theoretical demonstration that superconducting circuits composed on inductors, capacitors, and Josephson junctions are sufficient to create protected gates.

Our starting point is a fluxonium-like qubit with a high tunnel barrier. Half a flux quantum in the qubit loop produces a double well potential in phase across the small junction. With sufficient shunt capacitance, the two lowest energy states are approximately degenerate and can be described in a basis where they have either positive or negative circulating current. Such a qubit has highly biased noise sensitivity with respect to the Pauli operators. The tunnel barrier prevents tunneling between the two wells, causing bit flips (X noise) to be highly suppressed. On the other hand, flux noise in the loop tilts the double well potential and splits the ground states. We associate the Z basis eigenstates with the circulating current states. Thus flux noise in the loop acts like Z field noise.

\begin{figure}
	\includegraphics{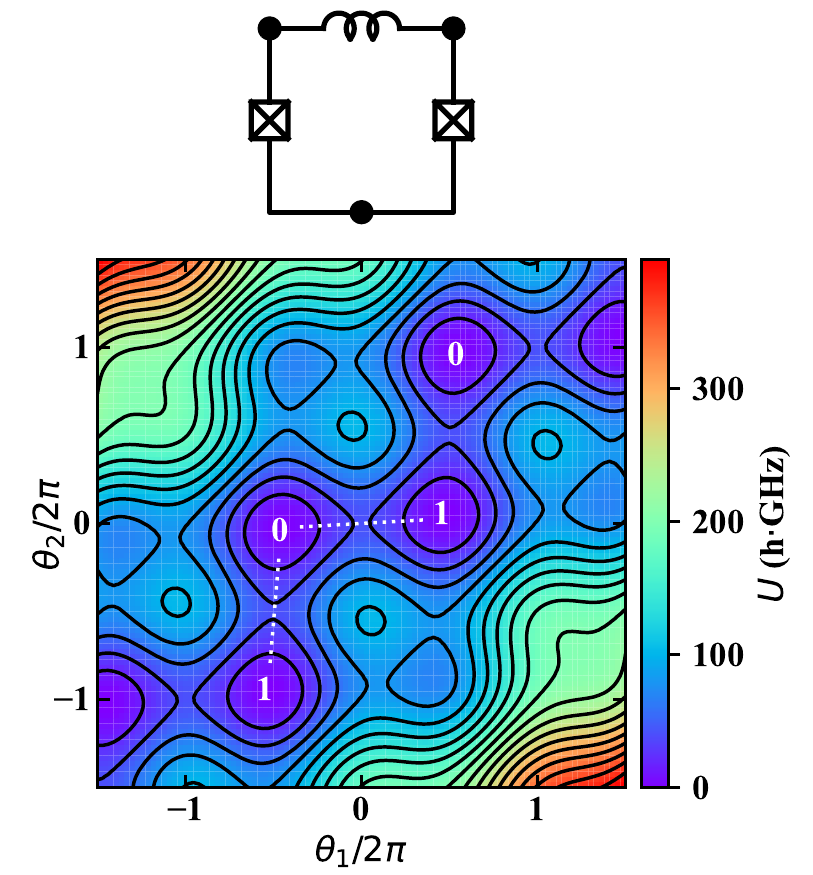}
	\caption{\label{fig:dual-barrier} Bifluxon circuit and potential. (top) The circuit is a single loop of two junctions and an inductor. (bottom) The potential $U$ as a function of the phase across each junction for half a flux quantum in the loop. The junctions have 50 nA critical current and the inductor has 80 nH inductance.  The potential shows two nonequivalent minima labeled 0 and 1 and two tunneling paths between them. The tunneling directions are nearly orthogonal, which makes them approximately independent.}
\end{figure}

\begin{table}[b]
\caption{\label{tab:table1} The frequency difference between the ground and first excited state of the dual-barrier fluxonium for various capacitance values. The superinductor has a 1~fF shunt and nodes have 1~fF to ground. }
\begin{ruledtabular}
\begin{tabular}{ccc}
\textrm{$C_{J1}$ (fF)}&
\textrm{$C_{J2}$ (fF)}&
\textrm{$f_{01}$(h$\cdot$GHz)}\\
\colrule
1.0 & 1.0 & 3.77\\
25.0 & 1.0 & 1.41\\
1.0 & 25.0 & 1.41\\
25.0 & 25.0 & $1.1\times10^{-4}$\\
\end{tabular}
\end{ruledtabular}
\end{table}

\begin{figure}
	\includegraphics{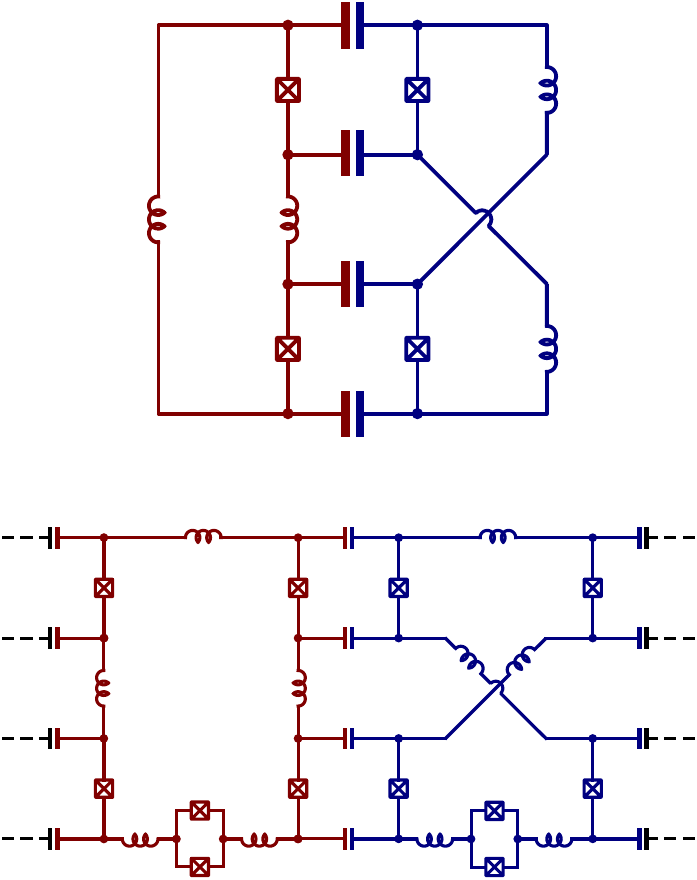}
	\caption{\label{fig:xx-circuits} (top) The basic XX-coupled dual-barrier fluxonium circuit. The circuit parameters for simulations are: 298~nA critical current, each junction; 10.1~nH, each inductor; 37.8~fF, each coupling capacitor; 0.017~fF to ground, each node; 0.084~fF shunt, each junction and inductor. (bottom) A variant that is extensible to longer chains with added compound junctions for tunable X fields.}
\end{figure}

\begin{figure*}
	\includegraphics{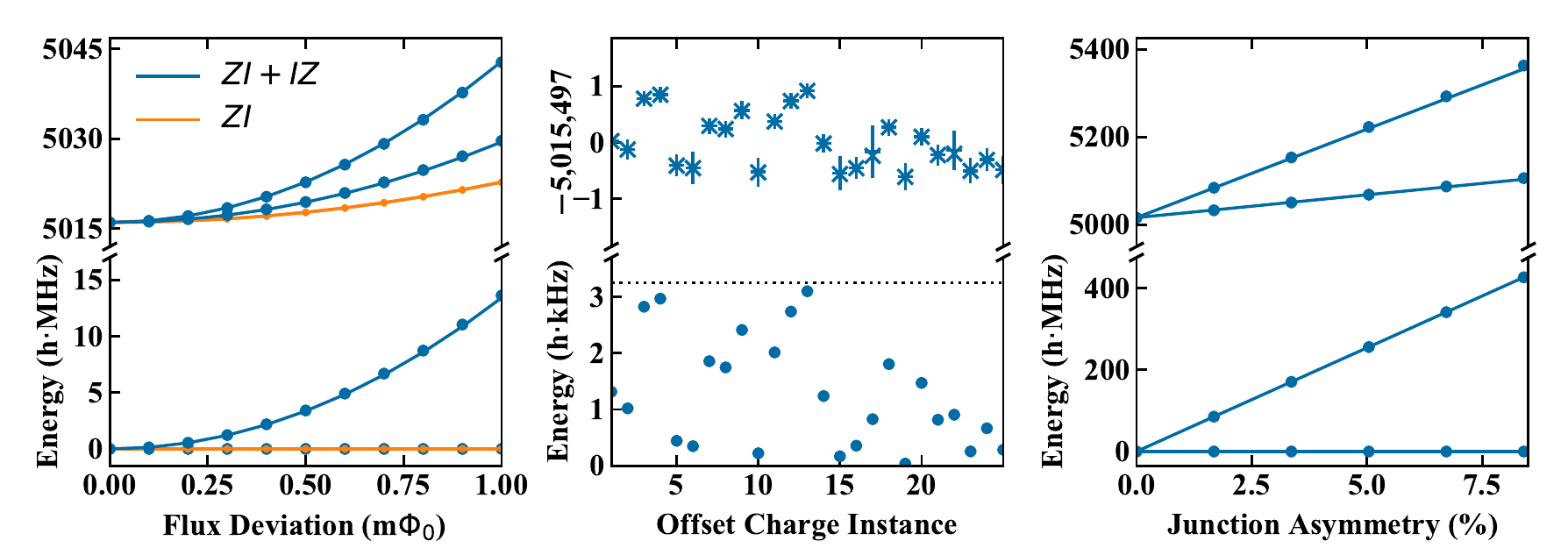}
		\caption{\label{fig:spec-vs-all} Numerical simulation of XX-coupled superconducting qubit circuit of Fig.~\ref{fig:xx-circuits}(top). (left) Spectrum versus flux deviation from $0.5 \Phi_0$ in the left qubit's loop (orange points) or both qubits' loops (blue circles). Solid lines are from fits to the Pauli models $E XX + s\Phi ZI$ or $E XX + s\Phi (ZI + IZ)$, where $E/h=5.015$~GHz and $s/h=129.5$~MHz/m$\Phi_0$ are the two fit parameters and $\Phi$ is the flux deviation from $0.5 \Phi_0$. (middle) Energy of the first three excited states relative to the ground state for 25 instances of random offset charge sampled uniformly from $0$ to $2e$ on each node. The dotted line marks the qubit energy splitting for zero offset charge on all nodes. The data above the break (``+'' and ``$\times$'' symbols) have roughly 5~GHz subtracted off, as indicated, to display their kHz spread. (right) Spectrum versus junction asymmetry on the left qubit. Junction asymmetry is $\delta \equiv 2(I_{c1}-I_{c2})/(I_{c1}+I_{c2})$. Solid lines are a fit to $E[(1+a\delta)(XX + YY - f ZZ) + (1-a\delta)(XX - YY + f ZZ)]/4$, with fit parameters $E/h=5.015$~GHz, $a=0.805$, and $f=0.258$. Estimated energy errors from basis truncation are given by error bars (middle) or are smaller than the data symbols.}
\end{figure*}

Creating a ZZ interaction is readily achieved via galvanic or magnetic mutual inductance between inductors. Tunable ZZ couplers between flux qubits are standard components in commercially available hardware \cite{harris}. We therefore focus on XX interactions, which are not as readily achieved. Our approach relies on controlling the co-tunneling of wavefunctions of a pair of flux qubits.

First, consider what would happen if one were to capacitively couple two fluxonium-like qubits as in Fig.~\ref{fig:fluxonia}. At half a flux quantum in both loops, the circuit produces a four-well potential as shown in the upper right plot. Minima are distinguished by the two circulating current directions in each loop, which we label 00, 01, 10, and 11. The coupling capacitors produce an anisotropic ``mass'' that elongates the wavefunctions along the 00 $\leftrightarrow$ 11 direction and squeezes the wavefunction in the 01 $\leftrightarrow$ 10 direction. The ground state is labeled $(|00\rangle + |11\rangle)/\sqrt{2}$. The first two excited states are nearly degenerate and labeled $(|01\rangle\pm|10\rangle)/\sqrt{2}$. The third excited state is $(|00\rangle - |11\rangle)/\sqrt{2}$. By fitting to the spectrum, we obtain the Pauli Hamiltonian $H = -XX + YY - 0.137ZZ$, where we have used the wavefunctions to identify the roles of X, Y and Z. Note that if the capacitors were crossed, the Hamiltonian would be $H = -XX - YY + 0.137ZZ$. Thus, if we could produce both couplings at the same time, we could cancel out the YY and ZZ terms. We cannot merely include straight and crossed coupling capacitors, however, because they interfere with each other, eliminating the coupling altogether.

A second piece to creating an XX interaction is the bifluxon \cite{bifluxon} circuit in Fig.~\ref{fig:dual-barrier}. By creating a loop with two small junctions and a large inductor, it is possible to make a double well potential with two nonequivalent tunneling paths between the two wells. With a sufficiently large inductance in the loop, a given tunneling path corresponds approximately to the phase slip of one junction and not the other.

To characterize the independence of the tunneling paths, we simulate the spectrum of the bifluxon circuit for different shunt capacitances across the junctions. The capacitive shunt is a proxy for capacitive coupling to other circuits. If each tunneling path is associated with the phase slip across an individual junction, then shunting one junction with a large capacitance should suppress tunneling along one path but not the other. Table~\ref{tab:table1} contains a summary of the tunnel-splitting between the ground and first excited state for various capacitive shunt arrangements: one large one small, two small, and two large capacitive shunts. Indeed, the data shows that the two tunneling paths are largely independent. This can be characterized by an efficiency for a given tunneling suppression. We calculate the efficiency as $\eta = E_{bs}E_{sb}/E_{ss}$, where $E_{sb}$ is the tunneling energy for one small ($s$) and one big ($b$) capacitor and $E_{ss}$ is for two small capacitors. The tunneling suppression is the value of $E_{bb}$. Now if the tunneling paths were completely independent, then the tunnel splittings (for $E_{sb} = E_{bs}$) would be exactly half of $E_{ss}$ and $\eta$ would equal 1. For the capacitances listed in Table~\ref{tab:table1}, the efficiency is $\eta=0.75$ with suppression of $10^{-4}$. Both efficiency and suppression depend on the choice of big and small capacitances.

Note that an isolated bifluxon circuit requires tuning of offset charge. We will show below that our XX coupled circuits eliminate offset charge sensitivity via large coupling capacitances.

\begin{figure}
	\includegraphics{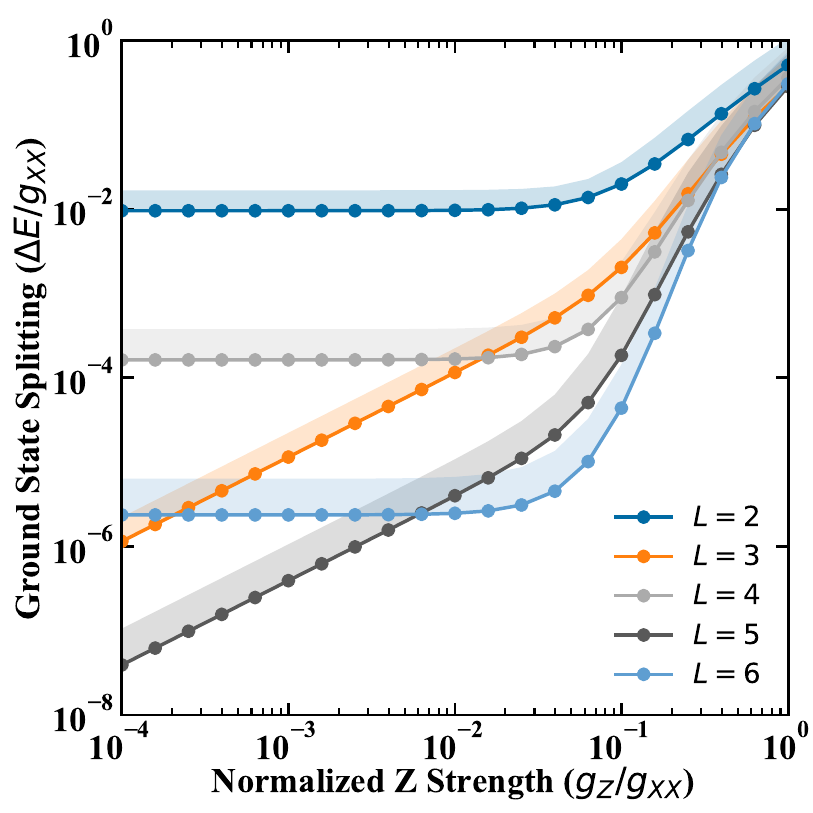}
	\caption{\label{fig:xxyy-effects} The effect of inadvertent YY and ZZ couplings. Average ground state splitting versus normalized Z field strength for a number of different chain lengths for 100 instances of Gaussian distributed YY and ZZ strengths.  Shaded regions extend to one standard deviation. Coupling terms are $-g_{XX}XX + |g_{YY}|YY - 0.2|g_{YY}|ZZ$. Plots are for $g_{YY}/g_{XX}$ with mean 0 and standard deviation of 1\%. Absolute values on couplings are included to avoid averaging out effects of YY and ZZ. (In separate simulations we observed that, for chains with mixed coupling signs, the splitting can decrease with increasing Z field over some range.)}
\end{figure}

We are now in a position to understand the full XX coupler of Fig.~\ref{fig:xx-circuits} (top). Bifluxon-like circuits exhibit two independent tunneling paths, while straight and crossed capacitive coupling cancel out YY and ZZ interactions. In the figure, the right qubit is twisted instead of the capacitors being crossed but the concept is the same. In addition to the basic circuit that we simulate below, we also show an extensible variant (bottom) which has four junctions per qubit loop. In addition, compound junctions are included to produce the tunable X fields necessary for the protected gates. We do not present simulations of this circuit but have verified that adding more junctions to a qubit loop creates more approximately independent tunneling paths.

We simulate the XX coupled qubit circuit with Circuitizer \footnote{Circuitizer contributors: R.~J. Epstein (creator and lead), D.~J. Clarke (co-lead), M.~E. Weippert, D.~G. Ferguson, E. Leonard Jr., R.~J. Magyar, K.~R. Colladay, W. DeGottardi, A. Setser, G.~R. Boyd, A. Marakov, K. W. Mahmud, A. Sirota}, an internally developed Python library which automates much of the process. Starting from a netlist containing circuit elements, their parameters, and their connectivity, the software generates a quantum Hamiltonian for diagonalization. Periodic and nonperiodic phase variables, and variables with conserved charges are automatically manifested via classical canonical coordinate transformations. Quantum harmonic oscillator eigenstates serve as the basis for nonperiodic variables whereas charge eigenstates are used for periodic ones. To achieve an estimated eigenvalue accuracy, the number of basis states for each degree of freedom is increased until eigenvalues change by less than the target accuracy. The 8-node circuit of Fig.~\ref{fig:xx-circuits} (top) has two conserved charges, two periodic variables, and two nonperiodic variables. See the Appendix for the circuit Hamiltonian generated by Circuitizer.

In Fig.~\ref{fig:spec-vs-all} (left) we plot the lowest energy levels as a function of flux deviation from $0.5\Phi_0$ in one or both qubit loops. This is equivalent to applying ZI or ZI + IZ to the XX qubit Hamiltonian. Two-parameter fits to the qubit Hamiltonian show very good agreement. The selected circuit parameters (see Fig.~\ref{fig:xx-circuits} caption) produce a 5-GHz gap, consistent with the gate simulations.

In the middle panel we plot the offset charge sensitivity of the circuit. Offset charges cause interference between the two tunneling paths but the effect is strongly suppressed by the capacitive coupling. Indeed the undesired tunneling strength is only 3.3 kHz, which sets a bound on the dephasing rate that offset charge can contribute. Nonzero offset charges can only decrease this tunnel splitting, which we confirm in the simulation for 25 instances of random offset charges. Likewise the offset charges change the 5-GHz gap by less than 2 kHz. Thus, the circuit has very low sensitivity to static offset charge. We have confirmed in separate simulations (not shown) that, like the transmon \cite{tmonkoch}, the residual tunnel-splitting is proportional to $\exp(-a(E_J/E_C)^{1/2})$, where $a$ is a constant, $E_J$ is the Josephson energy, and $E_C$ is the coupling capacitive energy. See the Appendix for parameters that achieve $5\times$ lower residual tunnel-splitting.

We also simulated the effects of junction asymmetry. The right panel shows how the nominally degenerate ground and excited states split linearly as the junctions of one qubit circuit deviate from their nominal value in opposite directions. The splitting is caused by the imperfect cancellation of the YY and ZZ couplings and the spectrum is well fit by the Pauli model. That the degeneracy is so sensitive to junction asymmetry seems concerning at first glance. However, as we show below, the encoded qubits are moderately tolerant of small YY and ZZ terms.

In Fig.~\ref{fig:xxyy-effects}, we plot the average ground state splitting due to Z fields for XX chains with small, random amounts of YY and ZZ couplings. The plot shows that YY and ZZ terms introduce splittings of the ground states at zero Z field or linear splitting sensitivity depending on whether the chain length is even or odd respectively. Importantly, both the zero-field splittings and linear sensitivities are strongly reduced with increasing chain length. In addition, the slopes of the curves at larger $g_Z/g_{XX}$ do get steeper with increasing chain length, showing the benefit of longer chains. However, the slopes are smaller than the case of perfect XX couplings (cf. Fig.~\ref{fig:chain}), meaning the exponent is now less than the chain length. Note that the apparent insensitivity to field strength for small $g_Z/g_{XX}$ is an artifact of the log-log plot for curves with vertical offsets. 

One difficulty associated with other protected qubits, like the $0-\pi$ qubit \cite{bkp, zeropi-ferg, zeropi-gros, zeropi-gyenis}, is that ever more extreme device parameters are needed to achieve ever greater protection. In our scheme, chain length adds another degree of freedom with which to optimize protection. For the present incarnation, however, the device parameters are quite challenging. The parasitic capacitances are one to two orders-of-magnitude below typical experimental values \cite{zeropi-gyenis}. Use of high plasma frequency junctions, small overlap capacitors, and released devices \cite{blochnium} could reduce parasitics considerably. While parasitics are a significant obstacle here and for the $0-\pi$ qubit, our circuit does not appear to suffer from low-frequency harmonic modes that plague $0-\pi$. Classical normal mode analysis (not shown) indicates that the lowest mode frequency of our circuit chains is only weakly dependent on chain length. Despite the device parameter challenges, the circuit design presented here is an existence proof that the lumped element model of superconducting circuits admits an implementation of noise-protected qubits and gates.

\section{Conclusion}
In summary, we have shown that protected qubits and wide-margin gates can be constructed from basic Hamiltonian terms, X, Z, XX and ZZ, and that wide-margin operation is enabled by transistor-like nonlinear signal transduction. We demonstrated that it is possible to create a \textsc{cnot} gate that preserves noise bias using coupled qubits, bypassing a no-go theorem \cite{guillaud}. We explored a superconducting circuit implementation of the protected qubits and gates and highlighted excellent agreement with a Pauli model. Finally, we provided evidence of moderate robustness to imperfect XX interactions caused by device asymmetry. In future work, it would be useful to generalize this approach to other gates including protected non-Clifford gates. It could also be fruitful to search for superconducting circuit designs or other physical incarnations that have less challenging design parameters. In a more speculative vein, the gates discussed here can be extended to 2-dimensional lattices of XX coupled qubits, which form phase self-correcting logical qubits. It would be intriguing to explore the feasibility of an architecture based on phase self-correction with active error correction solely for bit flips.

\begin{acknowledgments}
Thanks to D.~G. Ferguson, B. Eastin, W.~G. Brown, M.~E. Weippert, J.~T. Anderson and D. Bacon for stimulating discussions and encouragement. This research was supported by the U.S. Army Research Office under contract W911NF-17-C-0024. The content of this publication does not necessarily reflect the position or the policy of the Government, and no official endorsement should be inferred.
\end{acknowledgments}
\appendix*
\section{}
\textbf{CNOT Control Pulses.} Control pulses are constructed from Gaussian error functions. Middle pulses are proportional to $\{\mathrm{erf}(r(t - t_j + w/2)) + \mathrm{erf}(r(-t + t_j + w/2))\}/2$ where $t$ is time, $r = 35/T$, $w = 0.355T$, $T$ is the total gate duration, and $t_j$ is the pulse offset. The start and end pulses are turned on at the beginning and end of the gate, respectively, so they are constructed from a single error function. The offset is $t_j = jT/3$ for pulse indices starting at $j=0$. For the basic gate, $r$ is the same, the offset is $T/2$, and the pulse width is $w = 0.85T$.

\textbf{Circuit Hamiltonian.} The Hamiltonian generated by Circuitizer in units of h$\cdot$GHz for the XX coupled qubit circuit is:
\begin{flalign}
\begin{aligned}
   &118.258 \hat{I} + 115.427 \hat{n}^2_{c0} -0.408 \hat{n}_{c0}\hat{n}_{c1}+230.547 \hat{n}_{c0}\hat{n}_{c2}\\
    &+0.408 \hat{n}^2_{c1}+115.427 \hat{n}^2_{c2} -0.408 \hat{n}_{c2}\hat{n}_{c3}+0.408 \hat{n}^2_{c3}\\
    &+0.814 \hat{n}^2_{h4}+9.677 \hat{n}^2_{h5} +9.677 \hat{n}^2_{h6}+16.491 \hat{n}^2_{h7}\\
    &+0.814 \hat{\theta}^2_{h4}+9.677 \hat{\theta}^2_{h5}+9.677 \hat{\theta}^2_{h6}+16.491 \hat{\theta}^2_{h7}\\
    &+14.782 \hat{h}^{}_{c1}e^{-0.822j\hat\theta}_{h5}e^{-2.29j\hat\theta}_{h6}+14.782 \hat{h}^{\dagger}_{c1}e^{0.822j\hat\theta}_{h5}e^{2.29j\hat\theta}_{h6} \\
    &-14.782 \hat{h}^{\dagger}_{c1}e^{2.29j\hat\theta}_{h5}e^{-0.833j\hat\theta}_{h6}-14.782 \hat{h}^{}_{c1}e^{-2.29j\hat\theta}_{h5}e^{0.833j\hat\theta}_{h6}\\
    &+14.782 \hat{h}^{}_{c3}e^{0.833j\hat\theta}_{h5}e^{2.29j\hat\theta}_{h6}+14.782 \hat{h}^{\dagger}_{c3}e^{-0.833j\hat\theta}_{h5}e^{-2.29j\hat\theta}_{h6} \\
    &-14.782 \hat{h}^{\dagger}_{c3}e^{2.29j\hat\theta}_{h5}e^{-0.822j\hat\theta}_{h6}-14.782 \hat{h}^{}_{c3}e^{-2.29j\hat\theta}_{h5}e^{0.822j\hat\theta}_{h6}\nonumber
\end{aligned}
\end{flalign}

Subscript letters $c$ and $h$ stand for charge and harmonic bases respectively. Numerical subscripts identify the degree of freedom. The purely harmonic degrees of freedom (subscripts 4 an 7) are uncoupled from the rest of the system and are removed after checking that their energy quanta are above the energy gap. Degrees of freedom 0 and 2 have associated conserved charges and are also removed.

\textbf{Lower Residual Tunnel-Coupling.} As discussed, the circuit parameters listed in Fig.~\ref{fig:xx-circuits} achieve 5-GHz coupling with residual splitting of 3.3~kHz. One set of circuit parameters that gives rise to 5-GHz coupling, with residual ground state splitting of 0.7~kHz, is: 313~nA critical current, each junction; 9.58~nH, each inductor; 42.3~fF, each coupling capacitor; 0.016~fF to ground, each node; 0.08~fF shunt, each junction and inductor.

\bibliography{xxchains}

\end{document}